\documentclass[seceq,preprint]{ptptex}

\usepackage{graphicx}
\usepackage{amsfonts}
\usepackage{latexsym}
\usepackage{amssymb}
\usepackage{epsfig}
\usepackage{amsmath,amssymb}

\usepackage{graphicx}

\preprintnumber[3cm]{
OU-HET 695/2011, HGU-CAP 009\\PTPTeX ver.0.8\\ February, 2011}

\title{%
 Low scale thermal leptogenesis in neutrinophilic Higgs doublet models
}

\author{  
Naoyuki \textsc{Haba}${}^{(1)}$  and  
Osamu \textsc{Seto}${}^{(2)}$
}

\inst{    
${}^{(1)}$Department of Physics, Osaka University, Toyonaka, Osaka 560-0043, 
 Japan, \\
${}^{(2)}$Department of Architecture and Building Engineering,
 Hokkai-Gakuen University,
 Sapporo 062-8605, Japan
}

\abst{    
It is well-known that 
 leptogenesis in low energy scale is difficult 
 in the conventional Type-I seesaw mechanism
 with hierarchical right-handed neutrino masses. 
We show that in a class of two Higgs doublet model,
 where one Higgs doublet generates masses of quarks and charged leptons 
 whereas the other Higgs doublet with a tiny vacuum expectation value 
 generates neutrino Dirac masses, large Yukawa couplings lead to
 a large enough CP asymmetry of the right-handed neutrino decay.
Thermal leptogenesis suitably works at low energy scale as  
 keeping no enhancement of lepton number violating wash out effects. 
We will also point out that
 thermal leptogenesis works well 
 without confronting gravitino problem  
 in a supersymmetric neutrinophilic Higgs doublet
 model with gravity mediated supersymmetry breaking. 
Neutralino dark matter and baryon asymmetry generation by thermal leptogenesis
 are easily compatible in our setup. 
}

\begin{document}

\maketitle


\section{Introduction}

An origin of cosmological baryon asymmetry is one of 
 the prime open questions in particle physics as well as in cosmology.
Among various mechanisms of baryogenesis, 
 leptogenesis~\cite{FukugitaYanagida}
 is one of the most attractive idea because of its simplicity and 
 the connection to neutrino physics. 
Particularly, thermal leptogenesis requires
 only the thermal excitation of heavy right-handed Majorana neutrinos
 which generate 
 tiny neutrino masses via the seesaw mechanism~\cite{Type1seesaw} and 
 provides several implications for the light neutrino mass 
 spectrum~\cite{Buchmulleretal}.
The size of CP asymmetry in a right-handed neutrino decay is, roughly speaking,
 proportional to the mass of right-handed neutrino.
Thus,
 we obtain only insufficiently small CP violation for a lighter right-handed neutrino mass. 
That is the reason why it has been regarded that 
 leptogenesis in low energy scale is in general difficult in the 
 conventional Type I seesaw mechanism~\cite{LowerBound,Davidson:2002qv}. 

On the other hand, 
 in supersymmetric models with conserved R-parity to avoid rapid proton decay,
 thermal leptogenesis faces with ``gravitino problem''
 that the overproduction of gravitinos spoils the success of 
 Big Bang Nucleosynthesis (BBN)~\cite{GravitinoProblem}, 
 whereas the stable lightest supersymmetric particle (LSP) becomes dark matter candidate. 
In order not to overproduce gravitinos,
 the reheating temperature after inflation should not be so high
 that right-handed neutrinos can be thermally produced~\cite{GravitinoProblem2}.
In the framework of gravity mediated supersymmetry (SUSY) breaking, 
 a few solutions, 
 e.g., gravitino LSP with R-parity violation~\cite{Buchmuller:2007ui}, 
 very light axino LSP~\cite{Asaka:2000ew}
 and strongly degenerated right-handed neutrino masses~\cite{ResonantLeptogenesis}, 
 have been proposed.

Recently, 
 a new class of two Higgs doublet models (THDM)~\cite{2hdm} has been considered 
 in Refs.~\cite{Ma,Nandi,Ma:2006km,Davidson:2009ha,Logan:2010ag,
 HabaHirotsu}.
The motivation is as follows.
As mentioned above, seesaw mechanism naturally realizes tiny  
 masses of active neutrinos 
 through heavy particles coupled with 
 left-handed neutrinos. 
However, 
 those heavy particles are almost decoupled in the low-energy effective
 theory, few observations are expected in collider experiments. 
Then, some people consider a possibility of
 reduction of seesaw scale to TeV~\cite{TeVseesaw,Haba:2009sd}, 
 where effects of TeV scale right-handed neutrinos might be 
 observed in collider 
 experiments such as Large Hadron Collider (LHC) and
 International Linear Collider (ILC). 
However, they must introduce a fine-tuning in order to obtain 
 both tiny neutrino mass 
 and detectable left-right neutrino mixing 
 through which 
 experimental evidences can be discovered.
Other right-handed neutrino production processes in extended models
 by e.g.,
 $Z'$ exchange~\cite{Zprime} or Higgs/Higgsino decay~\cite{CerdenoSeto}
 also have been pointed out. 
Here, let us remind that
 Dirac masses of fermions are proportional to their Yukawa couplings
 as well as a vacuum expectation value (VEV) of the relevant Higgs field.
Hence, the smallness of a mass might be due to not a small Yukawa coupling
 but a small VEV of the Higgs field.
Such a situation is indeed realized in some THDM. 
For example, 
 in Type-II THDM with a large $\tan\beta$,
 the mass hierarchy between up-type quark and down-type quark 
 can be explained by the ratio of Higgs VEVs, and 
 when $\tan\beta \sim 40$,
 Yukawa couplings of top and bottom quark are
 same scale of order of unity~\cite{Hashimoto:2004xp}.
Similarly, there is a possibility that 
 smallness of the neutrino masses comparing to those of quarks and
 charged leptons is originating from an extra Higgs
 doublet with the tiny VEV.  
This idea is that 
 neutrino masses are much smaller than other fermions 
 because the origin of them comes from different VEV of different
 Higgs doublet, and then 
 we do not need  
 extremely tiny neutrino Yukawa couplings. 
Let us call this kind of 
 model~\cite{Ma,Nandi,Ma:2006km,Davidson:2009ha,Logan:2010ag,HabaHirotsu}
 neutrinophilic Higgs doublet model. 
Especially, in models in Refs.\cite{Ma, HabaHirotsu}, 
 tiny Majorana neutrino masses are obtained  
 through a TeV scale Type-I seesaw mechanism 
 without requiring tiny Yukawa couplings. 

Notice that neutrino Yukawa couplings in neutrinophilic Higgs doublet models
 do not need to be so small. 
This fact has significant implication to leptogenesis.
The CP violation of right-handed neutrino decay
 is proportional to neutrino Yukawa coupling squared.
We can obtain a large CP violation for such large neutrino Yukawa couplings.
This opens new possibility of low scale thermal leptogenesis. 
In this paper, we will show that  
 CP asymmetry is enhanced and thermal leptogenesis suitably works 
 in multi-Higgs models with a neutrinophilic Higgs doublet field,
 where  
 the tiny VEV of the neutrinophilic Higgs field has 
 equivalently larger neutrino Yukawa couplings, and 
 then TeV-scale seesaw works well. 
We will show that 
 the thermal leptogenesis suitably works at low energy scale as  
 avoiding enhancement of lepton number violating wash out effects. 
We will also point out that
 thermal leptogenesis in gravity mediated SUSY breaking works well 
 without confronting gravitino problem in a supersymmetric model.

\section{Neutrinophilic Higgs doublet models}

\subsection{Minimal neutrinophilic THDM }
\label{subsec:Minimal}

Let us show a two Higgs doublet model, 
 which we call neutrinophilic THDM model,
 originally 
 suggested in Ref.~\cite{Ma}. 
In the model, one additional Higgs doublet $\Phi_{\nu}$, which gives only neutrino Dirac masses, 
 besides the SM Higgs doublet $\Phi$ and a discrete $Z_2$-parity are introduced.
The $Z_2$-symmetry charges (and also lepton number) are assigned 
 as the following table. 
\begin{table}[h]
\centering
\begin{center}
\begin{tabular}{|l|c|c|} \hline
fields  &  $Z_{2}$-parity & lepton number \\ \hline\hline
SM Higgs doublet, $\Phi$  &  $+$ &  0 \\ \hline
new Higgs doublet, $\Phi_{\nu}$ 
 &  $-$ & 0 \\ \hline
right-handed neutrinos, $N$  &  $-$ & $1$ \\ \hline
others  &  $+$ & $\pm 1$: leptons, $0$: quarks \\ \hline
\end{tabular}
\end{center}
\end{table}
%
Under the discrete symmetry, 
 Yukawa interactions are given by 
\begin{eqnarray}
{\mathcal L}_{yukawa}=y^{u}\bar{Q}_L \Phi U_{R}
 +y^d \bar{Q}_{L}\tilde{\Phi}D_{R}+y^{l}\bar{L}\Phi E_{R} 
 +y^{\nu}\bar{L}\Phi_{\nu}N+ \frac{1}{2}M \bar{N^{c}}N
 +{\rm h.c.}\; 
\label{Yukawa:nuTHDM}
\end{eqnarray}
where 
 $\tilde{\Phi}=i\sigma_{2}\Phi^{\ast}$, and   
 we omit a generation index. 
$\Phi_\nu$ only couples with $N$ by the $Z_2$-parity so that 
 flavor changing neutral currents (FCNCs) 
 are suppressed. 
Quark and charged lepton sectors are the same as Type-I THDM, 
 but notice that  
 this neutrinophilic THDM is quite different from 
 conventional Type-I, II, X, Y THDMs~\cite{2hdm}.

The Higgs potential of the neutrinophilic THDM is given by 
\begin{align}
V^\text{THDM} 
&
= m_\Phi^2 \Phi^\dag \Phi + m_{\Phi_\nu}^2 \Phi_\nu^\dag \Phi_\nu
-m_3^2\left(\Phi^\dag \Phi_\nu+\Phi_\nu^\dag \Phi\right)
+\frac{\lambda_1}2(\Phi^\dag \Phi)^2
+\frac{\lambda_2}2(\Phi_\nu^\dag \Phi_\nu)^2\nonumber \\
&\qquad+\lambda_3(\Phi^\dag \Phi)(\Phi_\nu^\dag \Phi_\nu)
+\lambda_4(\Phi^\dag \Phi_\nu)(\Phi_\nu^\dag \Phi)
+\frac{\lambda_5}2\left[(\Phi^\dag \Phi_\nu)^2
+(\Phi_\nu^\dag \Phi)^2\right]. 
\label{Eq:HiggsPot}
\end{align}
The $Z_2$-symmetry is softly broken by $m_3^2$. 
Taking a parameter set, 
\begin{equation}
m_\Phi^2 < 0, ~~~ m_{\Phi_\nu}^2 > 0, ~~~ |m_{3}^2| \ll m_{\Phi_\nu}^2,
\end{equation}
we can obtain the VEV hierarchy of Higgs doublets,
\begin{equation}
v^2 \simeq \frac{-m_\Phi^2}{\lambda_1},
 ~~~ v_{\nu} \simeq \frac{-m_{3}^2 v}{ m_{\Phi_\nu}^2 + (\lambda_3 + \lambda_4 + \lambda_5 ) v^2} ,
\end{equation}
 where we have decomposed the SM Higgs doublet $\Phi$ and
 the extra Higgs doublet $\Phi_{\nu}$  as 
\begin{eqnarray}
\Phi =
\left(
\begin{array}{c}
v+ \frac{1}{\sqrt{2}}\phi^{0}\\
\phi^{-}
\end{array}
\right)
,\;\;
\Phi_{\nu}=
\left(
\begin{array}{c}
v_{\nu}+\frac{1}{\sqrt{2}}\phi^{0}_{\nu} \\
\phi^{-}_{\nu}
\end{array}
\right). 
\end{eqnarray}
When we take values of parameters as 
 $m_\Phi \sim 100$ GeV, 
 $m_{\Phi_\nu} \sim 1$ TeV, 
 and $|m_{3}^2| \sim 10$ GeV$^2$, 
 we can obtain $v_\nu \sim 1$ MeV. 
The smallness of $|m_{3}^2|$ is guaranteed by the 
 ``softly-broken'' $Z_2$-symmetry. 

For a very large $\tan \beta=v/v_{\nu} (\gg 1)$ limit we are interested in,
the five physical Higgs boson states and those masses are respectively given by
\begin{eqnarray}
 H^\pm \simeq \ [\phi_\nu^\pm] ,       && ~~~ m^2_{H^\pm} \simeq m_\nu^2 + \lambda_3 v^2 , \\ 
 A \simeq  {\rm Im} [\phi_{\nu}^0] ,   && ~~~ m^2_A \simeq m_\nu^2 +
  (\lambda_3 + \lambda_4+ \lambda_5) v^2 ,       \\
 h \simeq {\rm Re} [\phi^0] ,          && ~~~ m^2_{h} \simeq 2 \lambda_1 v^2  , \\ 
 H \simeq {\rm Re} [\phi_\nu^0] ,      && ~~~ m^2_{H} \simeq m_\nu^2 + (\lambda_3 + \lambda_4+\lambda_5) v^2 ,
\end{eqnarray}
 where negligible ${\cal O}(v_{\nu}^2)$ and ${\cal O}(m_3^2)$ corrections are omitted. 
Notice that 
 $\tan\beta$ is extremely large so that 
 the SM-like Higgs $h$ is almost originated from $\Phi$,
 while other physical Higgs particles, $H^\pm, H, A$, are almost
 originated from $\Phi_\nu$. 
Since $\Phi_\nu$ has Yukawa couplings only with neutrinos and 
 lepton doublets, 
 remarkable phenomenology can be expected which 
 is not observed in other THDMs. 
For example, lepton flavor violation (LFV) processes 
 and oblique corrections are 
 estimated in Ref.~\cite{Ma}, and  
 charged Higgs processes in collider experiments
 are discussed 
 in Refs.~\cite{Davidson:2009ha, Logan:2010ag}~\footnote{
The model deals with Dirac neutrino version
 in neutrinophilc THDM, but
 phenomenology of charged lepton has a similar region in part.}.

The neutrino masses including one-loop radiative corrections~\cite{Ma:2006km} are estimated as 
\begin{equation}
(m_\nu)_{ij} = \sum_k\frac{y^{\nu}_{ik} v_\nu y^{\nu T}{}_{kj} v_\nu}{M_k} + \sum_k {y^\nu_{ik} y^{\nu T}{}_{kj} M_{k} \over 16
 \pi^{2}} 
\left[ 
{m_R^{2} \over m_R^{2}-M_{k}^{2}} \ln {m_R^{2} \over M_{k}^{2}} - 
{m_I^{2} \over m_I^{2}-M_{k}^{2}} \ln {m_I^{2} \over M_{k}^{2}} \right],
\end{equation}
where $m_R$ and $m_I$ are the masses of $ {\rm Re} [\phi^{0}]$ and
 ${\rm Im} [\phi_\nu^{0}]$ respectively.  
It is easy to see the tree level contribution gives ${\cal O} (0.1)$ eV neutrino masses
 for $M_k \sim 1$ TeV, $v_\nu \sim 1$ MeV and $y^{\nu} = {\cal O}(1)$.
The one-loop contribution is induced for a nonvanishing $\lambda_5$.
When 
 $m_R^{2} - m_I^{2} = 2 \lambda_5 v^{2} \ll m_0^{2} = (m_R^{2} + m_I^{2})/2$, 
\begin{equation}
({m}_\nu)_{ij} = {\lambda_5 v^{2} \over 8 \pi^{2}} 
\sum_k {y^\nu_{ik} y^\nu_{jk} M_{k} \over m_0^{2} - M_{k}^{2}} \left[ 
1 - {M_{k}^{2} \over m_0^{2}-M_{k}^{2}} \ln {m_0^{2} \over M_{k}^{2}}
 \right],
\end{equation}
and it shows 
\begin{eqnarray}
&&({m}_\nu)_{ij} = {\lambda_5 v^{2} \over 8 \pi^{2}} 
\sum_k {y\nu_{ik} y^nu_{jk} \over M_{k}} \left[ 
\ln {M_{k}^{2} \over m_0^{2}} - 1 \right], \;\; (M_{k}^{2} \gg m_0^{2}), \\
&&
({m}_\nu)_{ij} = {\lambda_5 v^{2} \over 8 \pi^{2} m_0^{2}} 
\sum_k y^\nu_{ik} y^\nu_{jk} M_{k}, \;\; (m_0^{2} \gg M_{k}^{2}), \\
&&
({m}_\nu)_{ij} \simeq {\lambda_5 v^{2} \over 16 \pi^{2}} 
\sum_k {y^\nu_{ik} y^\nu_{jk} \over M_{k}}, \;\; (m_0^{2} \simeq M_{k}^{2}). 
\end{eqnarray}
Thus, when the masses of Higgs bosons (except for $h$) and right-handed neutrinos are
 ${\mathcal O}(1)$ TeV, 
 light neutrino mass scale of order ${\mathcal O}(0.1)$ eV is 
 induced with 
 $\lambda_5 \sim 10^{-4}$.
Thus, whether tree-level effect is larger than 
 loop-level effect or not
 is determined by 
 the magnitude of $\lambda_5$ (and $m_A, m_H$),
 which contribute one-loop diagram.

\subsection{A UV theory of neutrinophilic THDM }
\label{subsec:HabaHirotsu}

Here let us show a model in Ref.~\cite{HabaHirotsu} as 
 an example of UV theory of the 
 neutrinophilic THDM. 
This model is constructed by introducing
 one gauge singlet scalar field $S$, 
 which has a lepton number,
 and $Z_3$-symmetry shown 
 as the following table. 
\begin{table}[h]
\centering
\begin{center}
\begin{tabular}{|l|c|c|} \hline
fields  &  $Z_{3}$-charge & lepton number \\ \hline\hline
SM Higgs doublet, $\Phi$  &  1 & 0 \\ \hline
new Higgs doublet, $\Phi_{\nu}$ 
 &  $\omega^{2}$ & 0 \\ \hline
new Higgs singlet, $S$ 
 &  $\omega$  &
 $-2$ \\ \hline
right-handed neutrinos, $N$  &  $\omega$  & 1 \\ \hline
others  &  1  &  $\pm 1$: leptons, $0$: quarks \\ \hline
\end{tabular}
\end{center}
\end{table}
Under the discrete symmetry, 
 Yukawa interactions are given by 
\begin{eqnarray}
{\mathcal L}_{yukawa}=y^{u}\bar{Q}^{L}\Phi U_{R}
 +y^d \bar{Q}_{L}\tilde{\Phi}D_{R}+y^{l}\bar{L}\Phi E_{R} 
 +y^{\nu}\bar{L}\Phi_{\nu}N+\frac{1}{2}y^{N}S\bar{N^{c}}N +{\rm h.c.} . 
\label{22}
\end{eqnarray}
The Higgs potential can be written as
\begin{eqnarray}
V=&m_\Phi^{2}|\Phi|^{2}+m_{\Phi_{\nu}}^{2}|\Phi_{\nu}|^{2}-m_S^{2}|S|^{2}
 -\lambda S^{3}-\kappa S\Phi^{\dagger}\Phi_{\nu}\nonumber\\
     &+\frac{\lambda_{1}}{2}|\Phi|^{4}+\frac{\lambda_{2}}{2}|\Phi_{\nu}|^{4}+\lambda_{3}|\Phi|^{2}|\Phi_{\nu}|^{2}+\lambda_{4}|\Phi^{\dagger}\Phi_{\nu}|^{2}\nonumber\\
     &+\lambda_{S}|S|^4+
  \lambda_{\Phi}|S|^{2}|\Phi|^{2}+\lambda_{\Phi_{\nu}}|S|^{2}|\Phi_{\nu}|^{2} + {\it h.c.} .
\label{Potential:HabaHirotsu}
\end{eqnarray}
$Z_3$-symmetry forbids 
 dimension four operators, 
 $(\Phi^\dagger \Phi_{\nu})^{2}$, 
 $\Phi^\dagger \Phi_{\nu}|\Phi|^2$, 
 $\Phi^\dagger \Phi_{\nu}|\Phi_{\nu}|^2$,
 $S^4$, $S^2|S|^2$, $S^2|\Phi|^2$,
 $S^2|\Phi_{\nu}|^2$, 
 and dimension two or three operators, 
 $\Phi^\dagger \Phi_{\nu}$, 
 $S|\Phi|^{2}$, $S|\Phi_\nu|^{2}$. 
Although there might be introduced small soft breaking 
 terms such as $m_3'^2\Phi^\dagger \Phi_{\nu}$ to avoid 
 domain wall problem, 
 we omit them here, for simplicity. 
It has been shown that, with $\kappa \sim 1$ MeV, the desirable hierarchy of 
 VEVs 
\begin{eqnarray}
&& v_s \equiv \langle S \rangle \sim  1 \;\hbox{TeV},\;\;\;
  v \sim  100 \;\hbox{GeV}, \;\;\;
  v_{\nu} \sim  1 \;\hbox{MeV}, 
\end{eqnarray}
 and neutrino mass  
\begin{equation}
 m_\nu \simeq \frac{y^{\nu2}  v_{\nu}^{2}} {M_N } .
\end{equation}
 with Majorana mass of right-handed neutrino $M_N = y^N v_s$ 
 can be realized~\cite{HabaHirotsu}. 
This is so-called 
 Type-I seesaw mechanism in a TeV scale, when
 coefficients 
 $y^\nu$ and $y^N$ are assumed to be 
 of order one. 
The masses of scalar and pseudo-scalar mostly from $S$ are
 given by
\begin{eqnarray}
 m^{2}_{H_{S}} &=& m_3^{2}+2\lambda_{S} v_s^2 , \;\;\;\;\;\;
 m^{2}_{A_{S}} = 9 \lambda v_s ,
\end{eqnarray} 
 in the potential Eq.~(\ref{Potential:HabaHirotsu}) without CP violation. 
For parameter region with $v_s \gg 1$ TeV, 
 both scalar and pseudo-scalar are heavier than other particles.
After integrating out $S$, thanks to the $Z_3$-symmetry, 
 the model ends up with an effectively neutrinophilic THDM 
 with approximated $Z_2$-symmetry, $\Phi \to \Phi, \Phi_\nu \to -\Phi_\nu$. 
Comparing to the neutrinophilic THDM, 
 the value of $m_3^2$, which is a soft $Z_2$-symmetry breaking
 term, is expected to be 
 $\kappa v_s$. 
$\lambda_{5}$ is induced by integrating out $S$, which is 
 estimated as ${\mathcal O}(\kappa^2/m_S^2)\sim 10^{-12}$.
Thus, the neutrinophilic THDM has an approximate $Z_2$-symmetry. 

As for the neutrino mass induced from one-loop diagram \footnote{
We would like to thank J. Kubo and H. Sugiyama 
 for letting us notice this topic.  
}, 
 UV theory induces small $\lambda_5\sim 10^{-12}$  
 due to $Z_3$-symmetry,
 so that radiative induced neutrino mass from 
 one-loop diagram
 is estimated as $\lambda_5 v^2/(4\pi)^2M \sim 10^{-4}$
 eV. 
This can be negligible comparing to light neutrino mass
 which is induced from 
 tree level Type-I seesaw mechanism. 
The tree level neutrino mass is 
\begin{equation}
m_\nu^{tree}
 \sim {y_\nu^2 v_\nu^2 \over M}\sim 
{y_\nu^2 \kappa^2v^2 \over v_s^2M},
\end{equation}
where we input 
 $v_\nu \sim {\kappa v \over v_s}$.  
On the other hand, one-loop induced neutrino mass is 
 estimated as 
\begin{equation}
m_\nu^{loop} \sim {\lambda_5 y_\nu^2 \over 16\pi^2}{v^2 \over M}
\sim {y_\nu^2 \over 16\pi^2}{\kappa^2 v^2 \over M^2 M}.
\end{equation}
Putting $M\sim v_s$, 
\begin{equation}
{m_\nu^{loop} \over m_\nu^{tree}}\sim 
{1 \over 16\pi^2},
\end{equation} 
which shows loop induced neutrino mass
 is always smaller than tree level mass
 if UV theory is the model of Ref.~\cite{HabaHirotsu}.

\subsection{Supersymmetic extension of neutrinophilic Higgs doublet model}
\label{subsec:Super}

Now let us show the supersymmetric extension of the neutrinophilic
 Higgs doublet model. 
The supersymmetric extension is straightforward by
 extending its Higgs sector to be a four Higgs doublet model.
The superpotential is given by
\begin{eqnarray}
{\mathcal W}&=&y^{u}\bar{Q}^{L}H_u U_{R}
 +y^d \bar{Q}_{L}{H_d}D_{R}+y^{l}\bar{L}H_d E_{R} 
 +y^{\nu}\bar{L}H_{\nu}N+M {N^{}}^2 \nonumber \\
&& +\mu H_uH_d + \mu' H_\nu H_{\nu'}
+\rho H_u H_{\nu'} + \rho' H_\nu H_d,
\end{eqnarray}
where $H_u$ ($H_d$) is Higgs doublet which gives mass of 
 up- (down-) sector. 
$H_\nu$ gives neutrino Dirac mass and $H_{\nu'}$ does not 
 contribute to fermion masses. 
For the $Z_2$-parity, 
 $H_u, H_d$ are even, while $H_\nu, H_{\nu'}$ are odd. 
The $Z_2$-partity is softly broken by the $\rho$ and $\rho'$. 
We assume that 
 $|\mu|, |\mu'| \gg |\rho|, |\rho'|$, and 
 SUSY breaking soft squared masses can trigger 
 suitable electro-weak symmetry breaking. 
The Higgs potential is given by 
\begin{eqnarray}
 V &=& (|\mu|^2 +|\rho|^2) H_u^\dag H_u + (|\mu|^2+|\rho'|^2) H_d^\dag H_d 
      + (|\mu'|^2 +|\rho'|^2) H_{\nu}^\dag H_{\nu} + (|\mu'|^2+|\rho|^2) H_{\nu'}^\dag H_{\nu'}  \nonumber \\
  && + \frac{g_1^2}{2} \left( H_u^\dag \frac{1}{2} H_u - H_d^\dag\frac{1}{2} H_d 
     + H_{\nu}^\dag \frac{1}{2} H_{\nu} - H_{\nu'}^\dag \frac{1}{2}H_{\nu'} \right)^2  \nonumber \\
  && + \sum_a \frac{g_2^2}{2} \left( H_u^\dag \frac{\tau^a}{2} H_u + H_d^\dag\frac{\tau^a}{2} H_d 
     + H_{\nu}^\dag \frac{\tau^a}{2} H_{\nu} + H_{\nu'}^\dag \frac{\tau^a}{2}H_{\nu'} \right)^2  \nonumber \\
  && + m_{H_u}^2 H_u^\dag H_u  + m_{H_d}^2 H_d^\dag H_d 
     + m_{H_\nu}^2 H_{\nu}^\dag H_{\nu}+ m_{H_{\nu'}}^2 H_{\nu'}^\dag H_{\nu'} \nonumber \\
  && + B \mu H_u \cdot H_d + B' \mu' H_{\nu}\cdot H_{\nu'}
 + \hat{B} \rho H_u \cdot H_{\nu'} +
 \hat{B}' \rho' H_{\nu}\cdot H_{d}\nonumber\\
&& + \mu^* \rho H_d^\dag H_{\nu'}+\mu^* \rho' H_u^\dag H_{\nu}+
 \mu'^* \rho' H_{\nu'}^\dag H_{d}+\mu'^* \rho H_\nu^\dag H_{u}
 + {\it h.c.} ,
\end{eqnarray}
 where $\tau^a$ and dot represent a generator of $SU(2)$ and
 its anti-symmetric product respectively.
We assume Max.[$|\hat{B}\rho|, |\hat{B}'\rho'|, |\mu\rho|,
 |\mu'\rho|,|\mu\rho'|,|\mu'\rho'|$] $\sim {\mathcal O}(10)$
 GeV$^2$, which triggers VEV hierarchy between
 the SM Higgs doublet and neutrinophilic Higgs doublets. 
Notice that 
 quarks and charged lepton have small non-holomorphic Yukawa 
 couplings with $H_\nu$, through one-loop diagrams 
 associated with small mass parameters of $\hat{B}\rho,
 \hat{B}'\rho', \mu\rho, \mu'\rho, \mu\rho', \mu'\rho'$. 
This situation is quite different from non-SUSY model,
 where these couplings are extremely suppressed by 
 factor of $v_\nu/v$. 
As for the gauge coupling unification, 
 we must introduce extra particles, but 
 anyhow,  
 the supersymmetric extension of neutrinophilic Higgs doublet model 
 can be easily
 constructed as shown above.

\section{Leptogenesis}

\subsection{A brief overview of thermal leptogenesis}

In the seesaw model, the smallness of the neutrino masses can be 
 naturally explained by the small mixing 
 between left-handed neutrinos and 
 heavy right-handed Majorana neutrinos $N_i$. 
The basic part of the Lagrangian in the SM 
 with right-handed neutrinos is described as 
\begin{eqnarray}
{\cal L}_{N}^{\rm SM}=-y^{\nu}_{ij} 
\overline{l_{L,i}} \Phi N_j 
-\frac{1}{2} \sum_{i} M_i \overline{ N^c_i} N_i + h.c. , 
\label{SMnuYukawa}
\end{eqnarray} 
where $i,j=1,2,3$ denote the generation indices, 
 $h$ is the Yukawa coupling, 
 $l_L$ and $\Phi$ are the lepton and the Higgs doublets, 
 respectively, and 
 $M_i$ is the lepton-number-violating mass term  
 of the right-handed neutrino $N_i$ 
 (we are working on the basis of 
 the right-handed neutrino mass eigenstates). 
With this Yukawa couplings, the mass of left-handed neutrino is expressed by the well-known formula
\begin{equation}
m_{ij} = \sum_k \frac{y^{\nu}_{ik}v y^{\nu}{}^T_{kj}v}{M_k}. 
\end{equation}

The decay rate of the lightest right-handed neutrino is given by
\begin{eqnarray}
\Gamma_{N_1} = \sum_j\frac{y^{\nu}_{1j}{}^\dagger y^{\nu}_{j1}}{8\pi}M_1 = \frac{(y^{\nu}{}^\dagger y^{\nu})_{11}}{8\pi}M_1.
\end{eqnarray}
Comparing to the Friedmann equation for a spatially flat spacetime 
\begin{equation}
H^2 = \frac{1}{3 M_P^2}\rho ,
\label{FriedmannEq}
\end{equation}
 with the energy density of the radiation  
\begin{equation}
\rho = \frac{\pi^2}{30}g_*T^4 ,
\end{equation}
 where $g_*$ denotes 
 the effective degrees of freedom of relativistic particles and $M_P \simeq 2.4 \times 10^{18}$ GeV is the reduced Planck mass,
the condition of the out of equilibrium decay $\Gamma_{N_1} < \left.H\right|_{T=M_1}$ is rewritten as
\begin{eqnarray}
\tilde{m}_1 \equiv (y^{\nu}{}^\dagger y^{\nu})_{11} \frac{v^2}{M_1} < 
\frac{8\pi v^2}{M_1^2} \left.H\right|_{T=M_1} 
\equiv m_* 
\end{eqnarray}
 with $ m_* \simeq 1\times 10^{-3}$ eV and $v=174$ GeV.

In the case of the hierarchical mass spectrum for right-handed neutrinos, 
 the lepton asymmetry in the Universe is generated 
 dominantly by CP-violating out of equilibrium decay of 
 the lightest heavy neutrino, $N_1 \rightarrow l_L \Phi^*$ 
 and $ N_1 \rightarrow \overline{l_L} \Phi $. 
Then, its CP asymmetry is given by~\cite{FandG}
\begin{eqnarray}
\varepsilon &\equiv& 
\frac{\Gamma(N_1\rightarrow \Phi+\bar{l}_j)-\Gamma(N_1\rightarrow \Phi^*+l_j)}
{\Gamma(N_1\rightarrow \Phi+\bar{l}_j)+\Gamma(N_1\rightarrow \Phi^*+l_j)} \nonumber \\
 &\simeq&  -\frac{3}{8\pi}\frac{1}{(y^{\nu} y^{\nu}{}^{\dagger})_{11}}\sum_{i=2,3}
\textrm{Im}(y^{\nu}y^{\nu}{}^{\dagger})^2_{1i} \frac{M_1}{M_i}, \qquad \textrm{for} \quad M_i \gg M_1 .
\end{eqnarray}
Through the relations of the seesaw mechanism, 
 this can be roughly estimated as 
\begin{eqnarray}
\varepsilon 
 \simeq \frac{3}{8\pi}\frac{M_1 m_3}{v^2} \sin\delta 
 \simeq 10^{-6}\left(\frac{M_1}{10^{10}\textrm{GeV}}\right)
 \left(\frac{m_3}{0.05 \textrm{eV}}\right) \sin\delta,  
 \label{epsilon}
\end{eqnarray}
where $m_3$ is the heaviest light neutrino mass normalized by
 $0.05$ eV which is a preferred to account for atmospheric neutrino
 oscillation data~\cite{atm}. 

Using the above $\varepsilon$, 
 the resultant baryon asymmetry generated via thermal leptogenesis 
 is expressed as  
\begin{equation}
\frac{n_b}{s} \simeq  C \kappa \frac{\varepsilon}{g_*}  , 
 \label{b-sRatio}
\end{equation}
where $\left. g_*\right|_{T=M_1} \sim 100$ , 
 the so-called dilution (or efficiency) factor $ \kappa \leq {\cal O}(0.1) $ denotes
 the dilution by wash out processes,
 the coefficient
\begin{equation}
C = \frac{8 N_f + 4 N_H}{22 N_f + 13 N_H} ,
\label{C}
\end{equation}
 with $N_f$ and $N_H$ being the number of fermion generation and Higgs doublet~\cite{C}
 is the factor of the conversion from lepton to baryon asymmetry
 by the sphaleron~\cite{KRS}.
In order to obtain the observed baryon asymmetry 
 in our Universe $n_b/s \simeq 10^{-10}$~\cite{WMAP}, 
 the inequality $\varepsilon \gtrsim 10^{-7}$ is required.
This can be rewritten as $M_1 \gtrsim 10^9$ GeV,
 which is the so-called Davidson-Ibarra bound for models with
 hierarchical right-handed neutrino mass spectrum~\cite{LowerBound,Davidson:2002qv}.

\subsection{leptogenesis in neutrinophilic THDM }

Now
 we consider leptogenesis in the neutrinophilic THDM with the extra Higgs doublet $\Phi_{\nu}$
 described in Sec.~\ref{subsec:Minimal}.
The relevant interaction part of Lagrangian Eq.~(\ref{Yukawa:nuTHDM}) is expressed as 
\begin{eqnarray}
{\cal L}_{N}=-y^{\nu}_{ij} 
\overline{l_{L,i}} \Phi_{\nu} N_j 
-\frac{1}{2} \sum_{i} M_i \overline{ N^c_i} N_i + h.c. .
\label{Yukawa:nuTHDM(2)}
\end{eqnarray} 
The usual Higgs doublet $\Phi$ in Eq.~(\ref{SMnuYukawa}) is replaced by 
 new Higgs doublet 
 $\Phi_{\nu}$.
Again, we are working on the basis of 
 the right-handed neutrino mass eigenstates. 
Then, with these Yukawa couplings,
 the mass of left-handed neutrino is given by 
\begin{equation}
m_{ij} = \sum_k \frac{y^{\nu}_{ik}v_{\nu} y^{\nu}{}^T_{kj}v_{\nu}}{M_k}. 
\end{equation}
Thus, for a smaller VEV of $v_{\nu}$, a larger $y^{\nu}$ is required.

The Boltzmann equation for the lightest right-handed neutrino $N_1$, 
 which is denoted by $N$ here, is given by
\begin{eqnarray}
\dot{n}_N+3Hn_N 
&=& -\gamma (N\rightarrow L\Phi_{\nu}) - \gamma (N\rightarrow \bar{L}\Phi_{\nu}^*) 
\qquad\textrm{:decay}\nonumber\\
&& +\gamma (L\Phi_{\nu}\rightarrow N) + \gamma (\bar{L}\Phi_{\nu}^* \rightarrow N) 
\qquad \textrm{:inverse decay}\nonumber\\
&& -\gamma (N L \rightarrow A \Phi_{\nu})-\gamma ( N \Phi_{\nu} \rightarrow L A) 
 -\gamma (N \bar{L} \rightarrow A \Phi_{\nu}^* )-\gamma ( N \Phi_{\nu}^* \rightarrow \bar{L} A)  \nonumber\\
&& +\textrm{inverse  processes} \qquad \qquad \qquad \qquad  : \textrm{s-channel scattering} \nonumber\\
&& -\gamma (N L \rightarrow A \Phi_{\nu})-\gamma ( N \Phi_{\nu} \rightarrow L A) -\gamma ( N A \rightarrow L \Phi_{\nu})  \nonumber\\
&& -\gamma (N \bar{L} \rightarrow A \Phi_{\nu}^* )-\gamma ( N \Phi_{\nu}^* \rightarrow \bar{L} A) -\gamma ( N A \rightarrow \bar{L} \Phi_{\nu}^*)  \nonumber\\
&& +\textrm{inverse  processes} \qquad \qquad \qquad \qquad 
: \textrm{t-channel scattering} \nonumber\\
&& -\gamma (N N \rightarrow {\rm Final}) + \gamma ({\rm Final} \rightarrow N N) 
: {\rm annihilation} \nonumber\\ 
&=& -\Gamma_D (n_N-n_N^{eq})-\Gamma_{scat} (n_N-n_N^{eq}) 
 -\langle\sigma v(\rightarrow \Phi, \Phi_{\nu})\rangle  (n_N^2-n_N^{eq}{}^2) 
\label{Boltzman:N}
\end{eqnarray}
 where $\Phi, \Phi_{\nu}$ and $A$ denote the Higgs bosons, the neutrinophilic
 Higgs bosons and gauge bosons, respectively. 
Notice that 
 usual $\Delta L =1$ lepton number violating scattering processes
 involving top quark is absent in this model,
 because $\Phi_{\nu}$ has 
 neutrino Yukawa couplings.
Although the annihilation processes $(N N \rightarrow {\rm Final})$ is noted in Eq.~(\ref{Boltzman:N}),
 in practice, this is not relevant because 
 the coupling $y^{\nu}_{i1}$ is so small, as will be shown later,
 to satisfy
 the out of equilibrium decay condition.

The Boltzmann equation for the lepton asymmetry $L \equiv l-\bar{l}$ is given by
\begin{eqnarray}
 && \dot{n}_L+3H n_L  \nonumber \\
&=& \gamma(N\rightarrow l\Phi_{\nu}) - \gamma( \bar{N} \rightarrow \bar{l}\Phi_{\nu}^*) \nonumber \\
 && -\{ \gamma(l\Phi_{\nu}\rightarrow N) - \gamma( \bar{l}\Phi_{\nu}^* \rightarrow \bar{N}) \} \qquad\textrm{:decay and inverse decay} \nonumber \\
&& -\gamma ( l A \rightarrow N \Phi_{\nu} )+\gamma ( \bar{l} A \rightarrow \bar{N} \Phi_{\nu}^* )
 -\gamma (N l \rightarrow A \Phi_{\nu}) \nonumber \\ 
&& +\gamma ( \bar{N} \bar{l} \rightarrow A \Phi_{\nu}^*)  \quad\textrm{:s-channel $\Delta L=1$ scattering} \nonumber\\
&& -\gamma (N l \rightarrow A \Phi_{\nu})+\gamma (\bar{N} \bar{l} \rightarrow A \Phi_{\nu}^*)-\gamma (l A \rightarrow N \Phi_{\nu})+\gamma (\bar{l} A \rightarrow \bar{N} \Phi_{\nu}^*) \nonumber\\
&&  -\gamma ( l \Phi_{\nu} \rightarrow N A)+\gamma ( \bar{l} \Phi_{\nu}^* \rightarrow \bar{N} A)  \quad\textrm{:t-channel $\Delta L=1$ scattering} \nonumber \\
&& +\gamma( \bar{l}\bar{l} \rightarrow \Phi_{\nu}^*\Phi_{\nu}^*)-\gamma(ll\rightarrow \Phi_{\nu}\Phi_{\nu}) \nonumber \\ &&
  +2 \{ \gamma(\bar{l}\Phi_{\nu}^*\rightarrow l\Phi_{\nu})- \gamma(l\Phi_{\nu}\rightarrow \bar{l}\Phi_{\nu}^*) \} \quad\textrm{:t and s-channel $\Delta L=2$ scattering}  \nonumber  \\
&=& \varepsilon\Gamma_D(n_N-n_N^{eq}) - \Gamma_W n_L
\end{eqnarray}
where
\begin{equation}
\Gamma_W =
 \frac{1}{2}\frac{n_N^{eq}}{n_{\gamma}^{eq}}\Gamma_N + \frac{n_N}{n_N^{eq}}\Gamma_{\Delta L=1,t}
 + 2\Gamma_{\Delta L=1,s}+ 2\Gamma_{\Delta L=2}
\end{equation}
 is the wash-out rate.

The condition of the out of equilibrium decay is given as
\begin{eqnarray}
\tilde{m}_1 \equiv (y^{\nu}{}^\dagger y^{\nu})_{11}\frac{v_{\nu}^2}{M_1} < 
\frac{8\pi v_{\nu}^2}{M_1^2} \left.H\right|_{T=M_1} 
\equiv m_* \left(\frac{v_{\nu}}{v}\right)^2 
\end{eqnarray}
Notice that for $v_{\nu} \ll v$ the upper bound on $\tilde{m}_1$ becomes more stringent,
 which implies that the lightest left-handed neutrino mass is almost vanishing $m_1 \simeq 0$.
Alternatively the condition can be expressed as
\begin{equation}
 ( y^{\nu}{}^{\dagger} y^{\nu})_{11} < 8 \pi \sqrt{ \frac{\pi^2 g_*}{90} }\frac{M_1}{M_P} .
 \label{OoEqDecay} 
\end{equation}
Hence, for the TeV scale $M_1$,
 the value of $ (y^{\nu}{}^{\dagger} y^{\nu} )_{11}$ must be very small, 
 which can be realized by taking all $y^{\nu}_{i1}$ to be small.
Under such neutrino Yukawa couplings $y^{\nu}_{i1} \ll y^{\nu}_{i2}, y^{\nu}_{i3}$ and
 hierarchical right-handed neutrino mass spectrum, 
 the CP asymmetry, 
\begin{eqnarray}
\varepsilon 
&\simeq & -\frac{3}{8\pi}\frac{1}{(y^{\nu}{}^{\dagger}y^{\nu})_{11}}
\left(\textrm{Im}(y^{\nu}{}^{\dagger}y^{\nu})^2_{12} \frac{M_1}{M_2}
 + \textrm{Im}(y^{\nu}{}^{\dagger}y^{\nu})^2_{13} \frac{M_1}{M_3} \right)  \nonumber \\
& \simeq & -\frac{3}{8\pi}\frac{m_{\nu} M_1}{v_{\nu}^2} \sin\theta \nonumber \\
& \simeq & -\frac{3}{16\pi} 10^{-6} \left(\frac{0.1 {\rm GeV}}{v_{\nu}}\right)^2
  \left(\frac{M_1}{100 {\rm GeV}}\right)
  \left(\frac{m_{\nu}}{0.05 {\rm eV}}\right) \sin\theta , 
\label{CPasym}
\end{eqnarray}
 is significantly enhanced due to 
 large Yukawa couplings $y^{\nu}_{2i}$ and $y^{\nu}_{3i}$ as well as  
 the tiny Higgs VEV $v_{\nu}$ .
The thermal averaged interaction rate of $\Delta L =2$ scatterings is expressed as 
\begin{eqnarray}
 \Gamma^{(\Delta L =2)}
 = \frac{1}{n_{\gamma}}
  \frac{T}{32 \pi (2 \pi )^4}
 \int ds \sqrt{s} K_1\left(\frac{\sqrt{s}}{T} \right) \int \frac{d \cos\theta}{2}\sum \overline{ |{\cal M}|^2}
\end{eqnarray}
 with
\begin{eqnarray}
\sum \overline{|{\cal M}|^2}
 &=& 2 \overline{|{\cal M}_{\rm t}|^2} + 2 \overline{|{\cal M}_{\rm
 s}|^2} 
 \simeq \sum_{j,(\alpha, \beta)} 2 |y^{\nu}_{\alpha j} y^{\nu}_{\beta
  j}{}^{\dagger}|\frac{s}{M_{N_j}^2}, \quad  \textrm{ for} \quad s \ll
  M_j^2 \ .
\end{eqnarray}
The decoupling condition
\begin{eqnarray}
 \Gamma^{(\Delta L =2)}
   < \sqrt{\frac{\pi^2 g_*}{90}} \frac{T^2}{M_P}, 
\end{eqnarray}
 for $T < M_1$ is rewritten as
\begin{eqnarray}
 \sum_i \left(\sum_j \frac{  y^{\nu}_{ij} y^{\nu}_{ji}{}^{\dagger} v_{\nu}^2}{M_j}\right)^2
   < 32 \pi^3 \zeta(3) \sqrt{\frac{\pi^2 g_*}{90}} \frac{v_{\nu}^4}{T M_P} . 
 \label{L2DecouplingCondition} 
\end{eqnarray}
For lower $v_{\nu}$, $\Delta L =2$ wash out processes are more significant.
Inequality~(\ref{L2DecouplingCondition}) gives
 the lower bound on $v_{\nu}$  
 in order to avoid too strong wash out.

\begin{figure}
    \centerline{\includegraphics{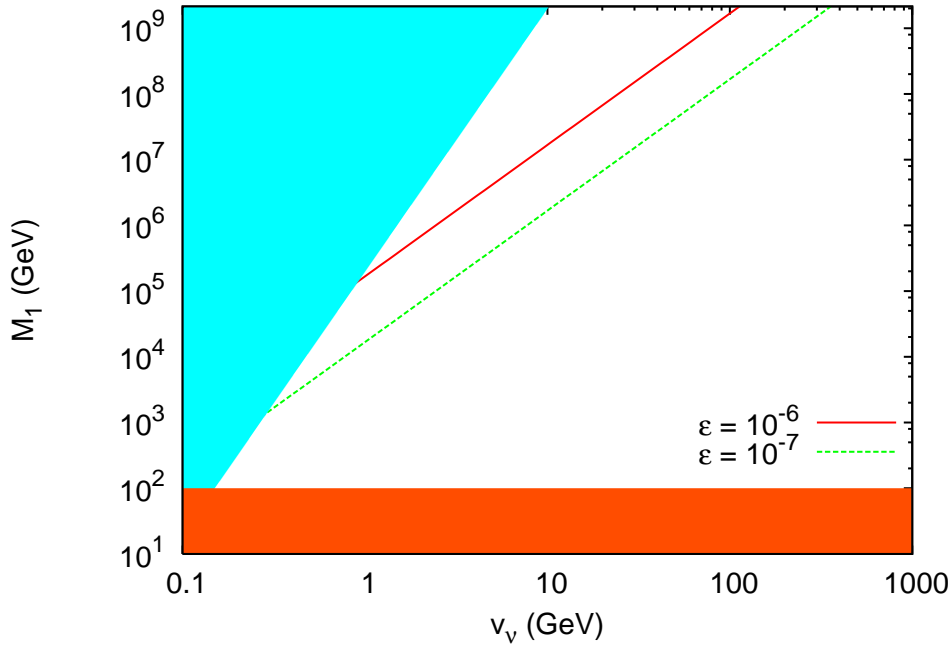}}
\caption{
 Available region for leptogenesis.
 The horizontal axis is the VEV of neutrino Higgs $v_{\nu}$ 
 and the vertical axis is the mass of the lightest right-handed neutrino mass $M_1$.
 In the red brown region, the lightest right-handed neutrino decay into Higgs boson $\Phi_{\nu}$ and lepton
 is kinematically forbidden.
 In turquoise region, $\Delta L=2$ wash out effect is too strong.
 The red and green line is contour of the CP asymmetry of $\varepsilon=10^{-6}$ and $10^{-7}$,
 respectively, with the lightest right-handed neutrino decay in 
 hierarchical right-handed neutrino mass spectrum.
}
\label{fig:AvailableRegion}
\end{figure}

We here summarize all conditions for successful thermal leptogenesis, 
 and the result is presented in Fig.~\ref{fig:AvailableRegion}.
The horizontal axis is the VEV of neutrino Higgs $v_{\nu}$ 
 and the vertical axis is the mass of the lightest right-handed
 neutrino, $M_1$.
In the red brown region, the lightest right-handed neutrino decay into Higgs boson $H$ with
 assuming $M_H= 100$ GeV, and lepton
 is kinematically not allowed.
In turquoise region corresponds to inequality~(\ref{L2DecouplingCondition}), $\Delta L=2$ wash out effect is too strong.
The red and green line is contour of the CP asymmetry of $\varepsilon=10^{-6}$ and $10^{-7}$,
 respectively, with
 the lightest right-handed neutrino decay in
 hierarchical right-handed neutrino mass spectrum.
Thus, in the parameter region above the line of $\varepsilon = 10^{-7}$, 
 thermal leptogenesis easily works 
 even with hierarchical masses of right-handed neutrinos.
For the region below the line of $\varepsilon = 10^{-7}$,
 the resonant leptogenesis mechanism~\cite{ResonantLeptogenesis}, where CP asymmetry is enhanced resonantly 
 by degenerate right-handed neutrino masses, may work.
Here we stress that, for $v_{\nu} \ll 100$ GeV, 
 the required degree of mass degeneracy is considerably milder than
 that for the original resonant leptogenesis. 

\subsection{Constraints on an UV theory}

Let us suppose that neutrinophilic THDM is derived from a model reviewed
 in Sec.~\ref{subsec:HabaHirotsu} by integrated out a singlet field $S$.
If $S$ is relatively light, thermal leptogenesis discussed above could be affected.
That is the annihilation processes of $N_1$ which has been justifiably ignored in Eq.~(\ref{Boltzman:N}).
However, the annihilation could take place more efficiently
 via S-channel $S$ scalar exchange processes in the UV theory~\cite{HabaHirotsu}.

For example, 
 the annihilation $N_1 N_1 \rightarrow \Phi_{\nu}\Phi_{\nu}^*$
 with the amplitude
\begin{eqnarray}
 \overline{ |{\cal M}|}^2
 = \left| \frac{y^{\nu}_1 \lambda_{\Phi_{\nu}} v_s}{ s - M_{H_S}^2 -i M_{H_S} \Gamma_{H_S}}\right|^2 \frac{s-4 M_1^2}{4} ,
\end{eqnarray}
 would not be in equilibrium, if 
\begin{eqnarray}
\lambda_{\Phi_{\nu}}
 \lesssim
 40 \frac{M_{H_S}^2}{M_1^{3/2} M_P^{1/2}}
 \simeq 0.1 \left(\frac{10^5 \, {\rm GeV}}{M_1}\right)^{3/2}\left(\frac{M_{H_S}}{10^7 \, {\rm GeV}}\right)^2 ,
\end{eqnarray}
 for $M_S \gg T > M_1$ is satisfied.
Here, $\Gamma_{H_S}$ denotes the decay width of $H_S$.
Constraints on other parameters such as $\lambda_\Phi$ and 
 $\kappa$ can be similarly obtained.

\section{Supersymetric case:
 Reconciling to thermal leptogenesis, gravitino problem and neutralino dark matter}

As we have shown in Sec.~\ref{subsec:Super}, it is possible to construct a supersymmetric model with $\Phi_{\nu}$. 
A discrete symmetry, called ``R-parity'', is imposed in many supersymmetric models
 in order to prohibit rapid proton decay.
Another advantage of the conserved R-parity is that it guarantees the absolute stability of
 the LSP, which becomes a dark matter candidate. 
In large parameter space of supergravity model with gravity mediated SUSY breaking,
 gravitino has the mass of ${\cal O}(100)$ GeV and decays into LSP (presumablly the lightest neutralino) 
 at late time after BBN. 
Then, decay products may affect the abundances of light elements produced during BBN.
This is so-called ``gravitino problem''~\cite{GravitinoProblem}.
To avoid this problem,
 the upper bound on the reheating temperature after inflation 
\begin{equation} 
 T_R < 10^6 - 10^7 \, {\rm GeV},
\label{ConstraintsOnTR}
\end{equation}
has been derived 
 as depending on gravitino mass~\cite{GravitinoProblem2}.
By comparing Eq.~(\ref{ConstraintsOnTR}) with
 the CP violation in supersymmetric models with hierarchical right-handed neutrino masses,
 which is about four times larger than that in non-supersymmetric
 model~\cite{SUSYFandG} as,
\begin{eqnarray}
\varepsilon 
 \simeq  -\frac{3}{2\pi}\frac{1}{(y^{\nu} y^{\nu}{}^{\dagger})_{11}}\sum_{i=2,3}
\textrm{Im}(y^{\nu}y^{\nu}{}^{\dagger})^2_{1i} \frac{M_1}{M_i},
\label{SUSYepsilon}
\end{eqnarray}
 it has been regarded that thermal leptogenesis
 through the decay of heavy right-handed neutrinos hardly work because of
 gravitino problem.

As we have shown in the previous section, 
 a sufficient CP violation $\varepsilon = {\cal O}(10^{-6})$ can be realized 
 for $v_{\nu} = {\cal O}(1)$ GeV 
 in the hierarchical right-handed neutrino masses with $M_1$ of ${\cal O}(10^5 - 10^6)$ GeV.
This implies
 that the reheating temperature after inflation $T_R$ of ${\cal O}(10^6)$ GeV is high enough 
 in order to produce right-handed neutrinos by thermal scatterings.
Thus, it is remarkable that 
 SUSY neutrinophilic model with $v_{\nu} = {\cal O}(1)$ GeV 
 can realize thermal leptogenesis
 in gravity mediated SUSY breaking with unstable gravitino.
In this setup, 
 the lightest neutralino could be LSP and
 dark matter with the standard thermal freeze out scenario.

\begin{figure}
    \centerline{\includegraphics
                                {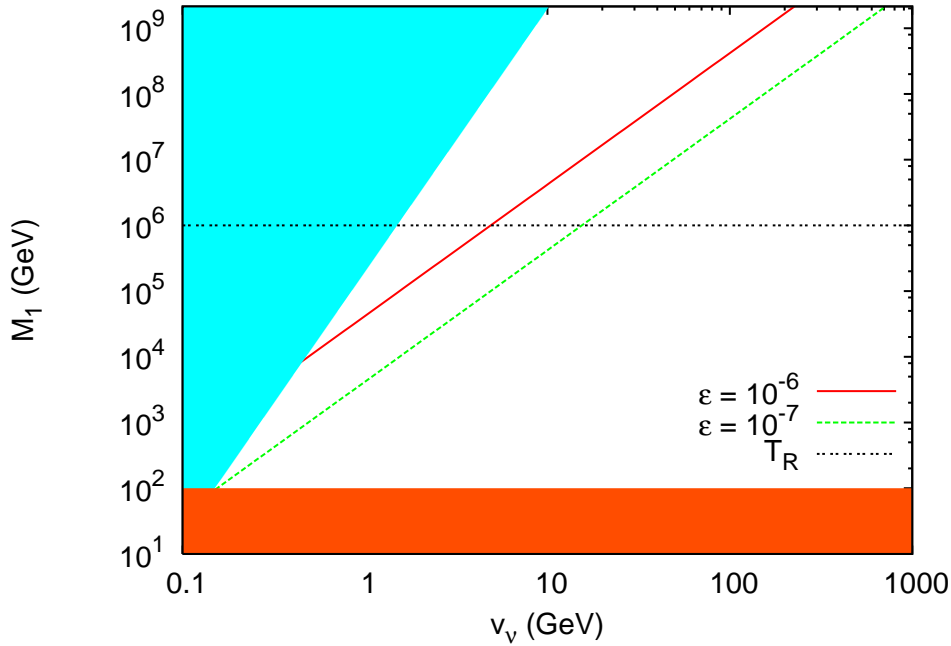}}
\caption{
 The same as Fig.~\ref{fig:AvailableRegion} but with Eq.~(\ref{SUSYepsilon}).
 The additional horizontal black dashed line represents a reference value of
 the upper bound on reheating temperature after inflation $T_R$ of $10^6$ GeV
 from gravitino overproduction.
 }
\label{fig:SUSYAvailableRegion}
\end{figure}

\section{Conclusion}

We have examined the possibility of thermal leptogenesis 
 in neutrinophilic Higgs doublet models, 
 whose tiny VEV gives neutrino Dirac mass term. 
Thanks to the tiny VEV of the neutrinophilic 
 Higgs field, 
 neutrino Yukawa couplings are not necessarily small, 
 instead, they tend to be large, and 
 the CP asymmetry in the lightest right-handed neutrino decay
 is significantly enlarged. 
Although 
 the $\Delta L = 2$ wash out effect also could be enhanced 
 simlitaneously,  
 we have found 
 the available parameter region where 
 its wash out effect is avoided as 
 keeping the CP asymmetry large enough. 
In addition, in a supersymmetric  
 neutrinophilic Higgs doublet model,  
 we have pointed out that
 thermal leptogenesis in gravity mediated SUSY breaking works well 
 without confronting gravitino problem. 
Where the lightest neutralino could be LSP and
 dark matter with the standard thermal freeze out scenario.

\section*{Acknowledgements}
We would like to thank M.~Hirotsu for collaboration
 in the early stage of this work.
We are grateful to S.~Matsumoto, S.~Kanemura and K.~Tsumura
 for useful and helpful discussions. 
This work is partially supported by Scientific Grant by Ministry of 
 Education and Science, Nos. 20540272, 22011005, 20039006, 20025004
 (N.H.), and
 the scientific research grants from Hokkai-Gakuen (O.S.).

\end{document}